\documentclass[10pt]{article}

\usepackage{amsmath,amssymb}

\usepackage{changepage}
\usepackage{arxiv}
\usepackage{textcomp,marvosym}

\usepackage{hyperref}

\usepackage[table]{xcolor}

\usepackage{array}

\usepackage{multirow}
\usepackage{lscape}
\usepackage{authblk}

\usepackage[sorting=nyt,style=numeric, natbib=true,doi=false,isbn=false,url=false,eprint=false]{biblatex}

\newcolumntype{+}{!{\vrule width 2pt}}

\newlength\savedwidth 

\newcommand\thickhline{\noalign{\global\savedwidth\arrayrulewidth\global\arrayrulewidth
2pt}%
\hline
\noalign{\global\arrayrulewidth\savedwidth}}

\usepackage[aboveskip=1pt,labelfont=bf,labelsep=period,justification=raggedright,singlelinecheck=off]{caption}

\makeatletter
\renewcommand{\@biblabel}[1]{\quad#1.}
\makeatother

\usepackage{graphicx}
\title{Good practices for clinical data warehouse
implementation: \\ a case study in France}
\author[1,2,*]{Matthieu Doutreligne}
\author[1]{Adeline Degremont}
\author[1]{Pierre-Alain Jachiet}
\author[3,4]{Antoine Lamer}
\author[5]{Xavier Tannier}

\affil[1]{Mission Data, Haute Autorité de Santé, Saint-Denis, France}
\affil[2]{Inria, Soda team, Palaiseau, France}
\affil[3]{Univ. Lille, CHU Lille, ULR 2694 - METRICS: Évaluation des
  Technologies de santé et des Pratiques médicales, Lille, France.}
\affil[4]{Fédération régionale de recherche en psychiatrie et santé mentale
  (F2RSM Psy), Hauts-de-France, Saint-André-Lez-Lille, France.}
\affil[5]{Sorbonne Université, Inserm, Université Sorbonne Paris-Nord,
  Laboratoire d'informatique médicale et d'ingénierie des connaissances en
  e-Santé, LIMICS, France}

\begin{document}
\maketitle

\bigskip

\begin{abstract}

  Real World Data (RWD) bears great promises to improve the quality of care.
  However, specific infrastructures and methodologies are required to derive
  robust knowledge and brings innovations to the patient.

  Drawing upon the national case study of the 32 French regional and university
  hospitals governance, we highlight key aspects of modern Clinical Data
  Warehouses (CDWs): governance, transparency, types of data, data reuse, technical
  tools, documentation and data quality control processes.

  Semi-structured interviews as well as a review of reported studies on French
  CDWs were conducted in a semi-structured manner from March
  to November 2022.

  Out of 32 regional and university hospitals in France, 14 have a CDW in
  production, 5 are experimenting, 5 have a prospective CDW project, 8 did not
  have any CDW project at the time of writing.  The implementation of CDW in
  France dates from 2011 and accelerated in the late 2020.

  From this case study, we draw some general guidelines for CDWs. The actual
  orientation of CDWs towards research requires efforts in governance
  stabilization, standardization of data schema and development in data quality
  and data documentation. Particular attention must be paid to the sustainability
  of the warehouse teams and to the multi-level governance. The transparency of
  the studies and the tools of transformation of the data must improve to allow
  successful multi-centric data reuses as well as innovations in routine care.
\end{abstract}


\clearpage

\section*{Introduction}\label{sec:introduction}

\subsection*{Real World Data}\label{subsec:rwd}

Health Information Systems (HIS) are increasingly collecting routine care data
\cite{jha_use_2009,sheikh_adoption_2014,kim_rate_2017,esdar_diffusion_2019,kanakubo_comparing_2019,liang_adoption_2021,apathy_decade_2021}.
This source of Real World Data (RWD) \cite{fda_real-world_2021} bears great
promises to improve the quality of care. On the one hand, the use of this data
translate into direct benefits --primary uses-- for the patient by serving as
the cornerstone of the developing personalized medicine
\cite{talukder_diseasomics_2022,mann_artificial_2022,ziegler_high_2022}. They also bring
indirect benefits --secondary uses-- by accelerating and improving knowledge
production: on pathologies \cite{campbell_characterizing_2022}, on the conditions of use of health products and
technologies\cite{safran_toward_2007,tuppin_value_2017}, on the measures of
their safety \cite{wisniewski_development_2003}, efficacy or usefulness in
everyday practice \cite{richesson_electronic_2013}. They can also be used to
assess the organizational impact of health products and technologies
\cite{has_guide_2020,has_real-world_2021}.

In recent years, health agencies in many countries have conducted extensive work
to better support the generation and use of real-life data
\cite{has_real-world_2021,kent_nice_2022,plamondongenevieve_integration_2022,fda_real-world_2021}.
Study programs have been launched by regulatory agencies: the DARWIN EU
program by the European Medicines Agency and the Real World Evidence Program by
the Food and Drug Administration \cite{fda_real_2018}.

\subsection*{Clinical Data Warehouse}

In practice, the possibility of mobilizing these routinely collected data
depends very much on their degree of concentration, in a gradient that goes from
centralization in a single, homogenous HIS to fragmentation in a multitude of
HIS with heterogeneous formats. The structure of the HIS reflects the governance
structure. Thus, the ease of working with these data depends heavily on the
organization of the healthcare actors.

Healthcare actors are sometimes concentrated in a small number of
organizations, resulting in uniform sources of real-life
data. For example, in Israel, the largest provider of provider (Clalit) insures
and cares for more than half of the population. In South Korea, the government
agency responsible for healthcare system performance and quality (HIRA) is
connected to the HIS of all healthcare stakeholders. England has a centralized
health care system under the National Health Service. This organization has
enabled it to bring together data from urban medicine in two large databases
databases that correspond to the two major software publishers. Currently,
opensafely \cite{opensafely_2022}, a first operating platform for research on Covid-19 exists and
should be followed by other similar platforms for more general themes.

Conversely, the production of real-life data may be distributed among many
entities, that have made different choices, without common management. Despite
heterogeneous insurance systems and hospitals in the United States, the grouping
of insurers into large entities nevertheless makes it possible to create large
databases such as Medicare, Medicaid or IBM MarketScan. Germany has found that
its data collection systems are very heterogeneous, limiting the potential of
health data. At the Medical Informatics Initiative \cite{gehring_german_2018},
it created four consortia in 2018 to develop technical and organizational
technical and organizational solutions to improve the consistency of clinical
data.

In France, the national insurer collects all hospital activity and city care
claims into a unique reimbursement database. However, clinical data is scattered
at each care site in numerous HISs.

\begin{figure*}
  \centering
  \includegraphics[width=0.7\linewidth]{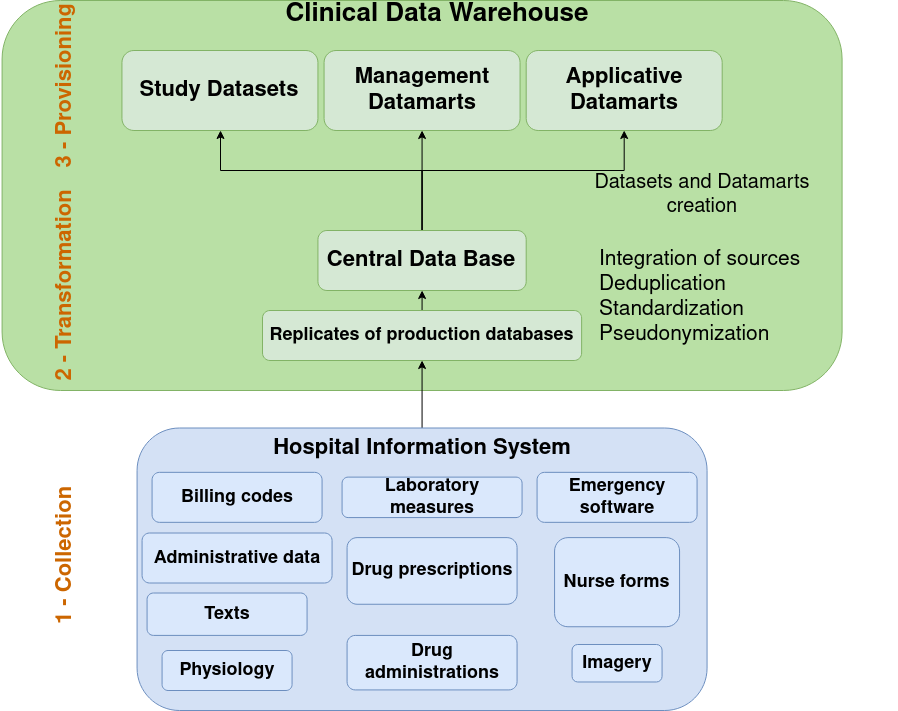}
  \caption{Clinical Data Warehouse: Three steps of data flow from the Hospital Information System: 1) collection, 2) transformations and 3) provisioning.}
  \label{background:CDW:fig:ehr_flow}
\end{figure*}

Whatever the organizational framework, it requires an infrastructure that pools
data from one or more medical information systems, to homogeneous
formats for management, research or care reuses
\cite{chute_enterprise_2010,pavlenko_implementation_2020}. Figure
\ref{background:CDW:fig:ehr_flow} illustrates for a Clinical Data Warehouse, the
three phases of data flow from the various sources that make up the HIS:
\begin{enumerate}
  \item \textbf{Collection} and copying of original sources.
  \item \textbf{Transformation}: Integration and harmonization
        \begin{itemize}
          \item Integration of sources into a unique database.
          \item Deduplication of identifiers.
          \item Standardization: A unique data model, independent of the
                software models harmonizes the different sources in a common schema,
                possibly with common nomenclatures.
          \item Pseudonymization: Removal of directly identifying elements.
        \end{itemize}
  \item \textbf{Provision} of sub-population data sets and transformed datamarts
        for primary and secondary reuse.
\end{enumerate}

In France, several hospitals deployed efforts for about ten years to create
CDWs from electronic medical records
\cite{cuggia_roogle_2011,jannot_georges_2017,garcelon_finding_2017,wack_installation_2017,daniel_initializing_2018,malafaye_mise_2018,artemova_predimed_2019,lelong_building_2019,conan_les_2021,
  lamer_development_2022}. This work has accelerated recently, with the beginning
of CDWs structuring at the regional and national levels. Regional cooperation
networks are being set up --such as the Ouest Data Hub \cite{hugo_2022}. In July
2022, the Ministry of Health opened a 50 million euros call for projects to set
up and strengthen a network of hospital CDWs coordinated with the national
platform, the Health Data Hub by 2025.

\subsection*{Objective}\label{objective}
Based on an overview of university hospital CDWs in France, this study make
general recommendations for properly leveraging the potential of CDWs to improve
healthcare. It focuses on: governance, transparency, types of
data, data reuse, technical tools, documentation and data quality control
processes.

\section*{Material and methods}\label{methods}

Interviews were conducted from March to November 2022 with 32 French regional
and university hospitals, both with existing and prospective CDWs.

\subsection*{Interviews}\label{methods:interviews}

Semi-structured interviews were conducted on the following
themes: the initiation and construction of the CDWs; the current status of the
project and the studies carried out; opportunities and obstacles; quality
criteria for observational research. Appendix \ref{apd:table:expert_teams} lists
all interviewed people with their team title. The complete form, with the
precised questions, is available in Appendix \ref{apd:interview_form}.

The interview form was sent to participants in advance, and then used as a
support to conduct the interviews. The interviews lasted 90 minutes and were
recorded for reference.

\subsection*{Quantitative methods}\label{methods:quantitative}

Three tables detailed the structured answers in Appendix \ref{apd:study_tables}.
The first two tables deal with the characteristics of the actors, and those of
the data warehouses. We completed them based on the notes taken during the
interviews, the recordings, and by asking the participants for additional
information. The third table focuses on ongoing studies in the CDWs. We
collected the list of these studies from the dedicated reporting portals, which
we found for 8 out of 14 operational CDWs. We developed a classification of
studies, based on the typology of retrospective studies described by the OHDSI
research network \cite{schuemie_book_2021}. We enriched this typology by
comparing it with the collected study resulting in the six following categories:

\begin{itemize}
  \item \textbf{Outcome frequency}: Incidence or prevalence estimation for a
        medically well-defined target population.
  \item \textbf{Population characterization}: Characterization of a specific set
        of covariates. Feasibility and pre-screening studies belong to this category \cite{pasco_pre-screening_2019}.
  \item \textbf{Risk factors}: Identification of covariates most associated with
        a well-defined clinical target (disease course, care event). These
        studies look at  association study without quantifying the causal effect
        of the factors on the outcome of interest.
  \item \textbf{Treatment Effect}: Evaluation of the effect of a well-defined
        intervention on a specific outcome target. These studies intend to show
        a causal link between these two variables \cite{hernan_methods_2021}.
  \item \textbf{Development of decision algorithms}: Improve or automate a
        diagnostic or prognostic process, based on clinical data from a given
        patient. This can take the form of a risk, a preventive score, or the
        implementation of a diagnostic assistance system. These studies are part
        of the individualized medicine approach, with the goal of inferring
        relevant information at the level of individual patient's files.
  \item \textbf{Medical informatics}: Methodological or tool oriented. These
        studies aim to improve the understanding and capacity for action of
        researchers and clinicians. This type of study includes the evaluation
        of a decision support tool, the extraction of information from
        unstructured data, automatic phenotyping methods.
\end{itemize}

Studies were classified according to this nomenclature based on their title and
description.


\section*{Results}\label{results}

\begin{figure*}[!b]
  \centering
  \includegraphics[width=0.67\linewidth]{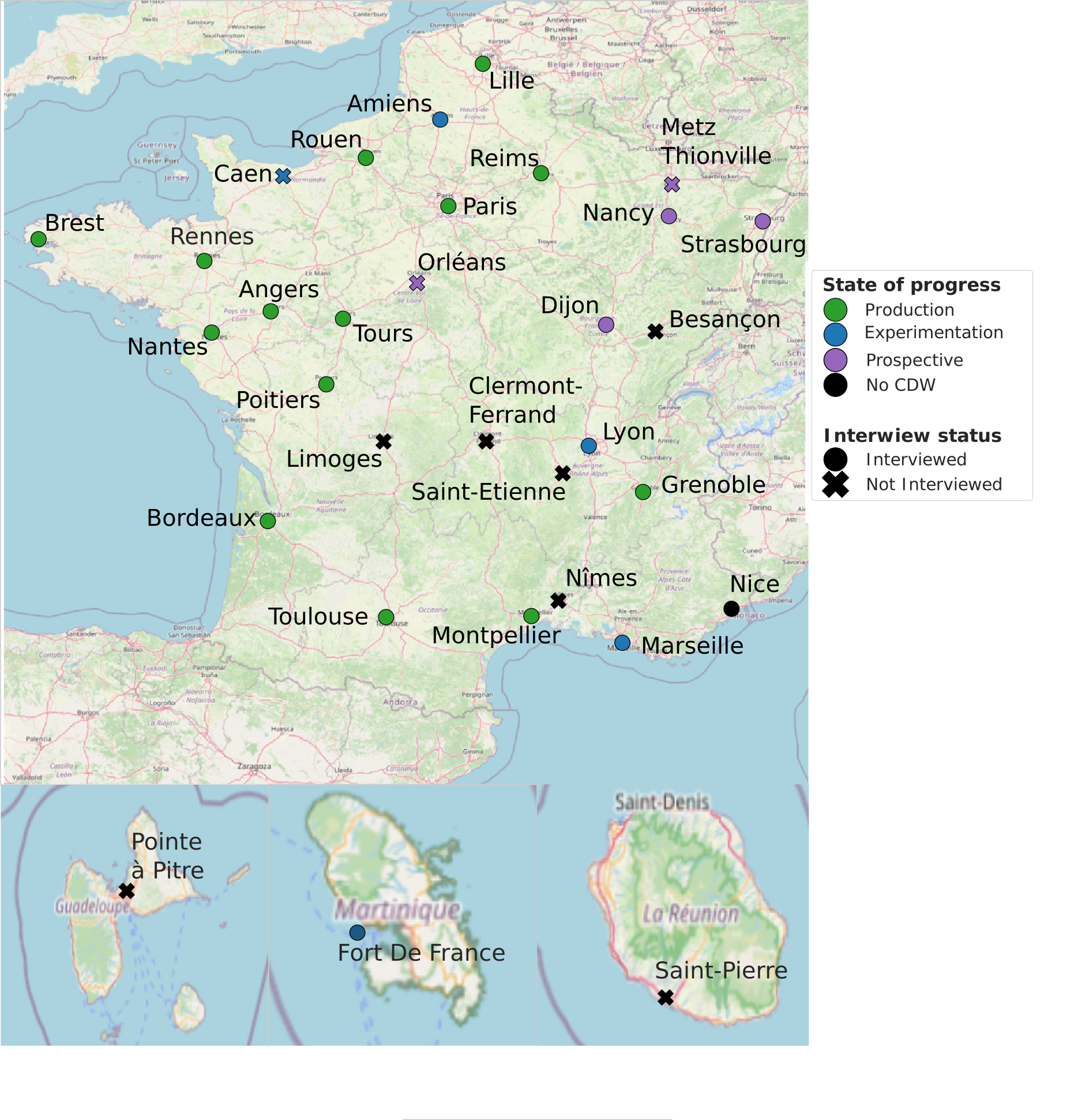}
  \caption{Repartition of CDWs in France.}
  \label{results:image:eds_map}
\end{figure*}

Figure \ref{results:image:eds_map} summarizes the development state of progress
of CDWs in France. Out of 32 regional and university hospitals in France, 14
have a CDW in production, 5 are experimenting, 5 have a prospective CDW project,
8 did not have any CDW project at the time of writing. The results are described
for all projects that are at least in the prospective stage minus the three that
we were unable to interview after multiple reminders (Orléans, Metz and Caen),
resulting in a denominator of 21 university hospitals.

\subsection*{Governance}

\begin{figure*}
  \centering
  \includegraphics[width=0.9\linewidth]{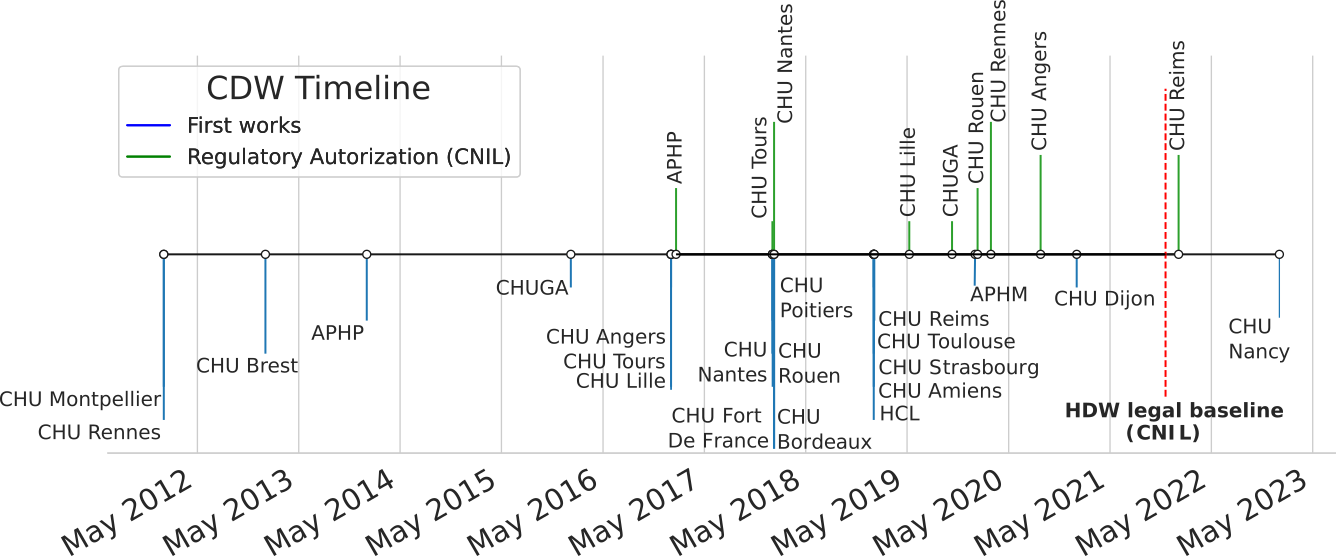}
  \hspace{5em}
  \caption{The French CDWs implementations date back to the first academic works in data reuse in early 2010s and accelerated recently.}
  \label{results:governance:image:timeline}
\end{figure*}

Figure \ref{results:governance:image:timeline} shows the history of the
implementation of CDWs. A distinction must be made between the first works --in
blue--, which systematically precede the regulatory authorization --in green--
from the French Commission on Information Technology and Liberties (CNIL).

The CDWs have so far been initiated by one or two people from the hospital world
with an academic background in bioinformatics, medical informatics or
statistics. The sustainability of the CDW is accompanied by the construction of
a cooperative environment between different actors: Medical Information
Department (MID), Information Systems Department (IT), Clinical Research
Department (CRD), clinical users, and the support of the management or the
Institutional Medical Committee. It is also accompanied by the creation of a
team, or entity, dedicated to the maintenance and operationalization of the CDW.
More recent initiatives, such as those of the HCL (Hospitals of the city of
Lyon) or the \textit{Grand-Est} region, are distinguished by an initial,
institutional and high-level support.

The CDW has a federating potential for the different business departments of the
hospital with the active participation of the CRD, the IT Department and the
MID. Although there is always an operational CDW team, the human resources
allocated to it vary greatly: from half a full-time equivalent to 80 people for
the AP-HP, with a median of 6.0 people. The team systematically includes a
coordinating physician. It is multidisciplinary with skills in public health,
medical informatics, informatics (web service, database, network,
infrastructure), data engineering and statistics.

Historically, the first CDWs were based on in-house solution development. More
recently, private actors are offering their services for the implementation and
operationalization of CDWs (15/21). These services range from technical
expertise in order to build up the data flows and clean them up to the delivery
of a platform integrating the different stages of data processing.

\subsection*{Management of studies}

Before starting, projects are systematically analyzed by a scientific and
ethical committee. A local submission and follow-up platform is often mentioned
(12/21), but its functional scope is not well defined. It ranges from simple
authorization of the project to the automatic provision of data into a Trusted
Research Environment (TRE) \cite{goldacre_better_2022}. The processes
for starting a new project on the CDW are always communicated internally but rarely documented publicly (8/21).

\subsection*{Transparency}

Studies underway in CDWs are unevenly referenced publicly on hospital websites.
In total, we found 8 of these portals out of 14 CDWs in production. Uses other
than ongoing scientific studies are very rarely publicly documented.

\subsection*{Data}

\subsubsection*{Strong dependance to the HIS}

CDW data reflect the HIS used on a daily basis by hospital staff. Stakeholders
point out that the quality of CDW data and the amount of work required for rapid
and efficient reuse are highly dependent on the source HIS. The possibility of
accessing data from an HIS in a structured and standardized format greatly
simplifies its integration into the CDW and then its reuse.

\subsubsection*{Categories of data}

Although the software landscape is varied across the country, the main
functionalities of HIS are the same. We can therefore conduct an analysis of the
content of the CDWs, according to the main categories of common data present in
the HIS.

The common base for all CDWs is constituted by data from the Patient
Administrative Management software (patient identification, hospital movements)
and from billing codes. Then, data flows are progressively developed from the
various softwares that make up the HIS. The goal is to build a homogeneous data
schema, linking the sources together, controlled by the CDW team. The
prioritization of sources is done through thematic projects, which feed the CDW
construction process. These projects improve the understanding of the sources
involved, by confronting the CDW team with the quality issues present in the
data.

Table \ref{results:data:img:data_categories} presents the different ratio of
data categories integrated in French CDWs. Structured biology and texts are
almost always integrated (20/21 and 20/21). The texts contain a large amount of
information. They constitute unstructured data and are therefore more difficult
to use than structured tables. Other integrated sources are the hospital drug
circuit (prescriptions and administration, 16/21), Intense Care Unit (ICU, 2/21)
or nurse forms (4/21). Imaging is rarely integrated (4/21), notably for reasons
of volume. Genomic data are well identified, but never integrated, even though
they are sometimes considered important and included in the CDW work program.


\begin{table}[!ht]
  \centering
  \begin{tabular}{lrl}
    \thickhline
    Category of data     & Number of CDW & Ratio  \\
    \thickhline
    Administrative       & 21            & 100 \% \\
    Billing Codes        & 20            & 95 \%  \\
    Biology              & 20            & 95 \%  \\
    Texts                & 20            & 95 \%  \\
    Drugs                & 16            & 76 \%  \\
    Imagery              & 4             & 19 \%  \\
    Nurse Forms          & 4             & 19 \%  \\
    Anatomical pathology & 3             & 14 \%  \\
    ICU                  & 2             & 10 \%  \\
    Medical devices      & 2             & 10 \%  \\
    \thickhline
  \end{tabular}
  \vspace{1em}
  \caption{Type of data integrated into the French CDWs: Text, billing codes and biology are the foundations that enrich the core administrative data.}\label{results:data:img:data_categories}
\end{table}

\subsection*{Data reuse}

\subsubsection*{Today, the main use put forward for the constitution of CDWs is that of scientific research.}

\begin{figure*}
  \centering
  \includegraphics[width=0.7\linewidth]{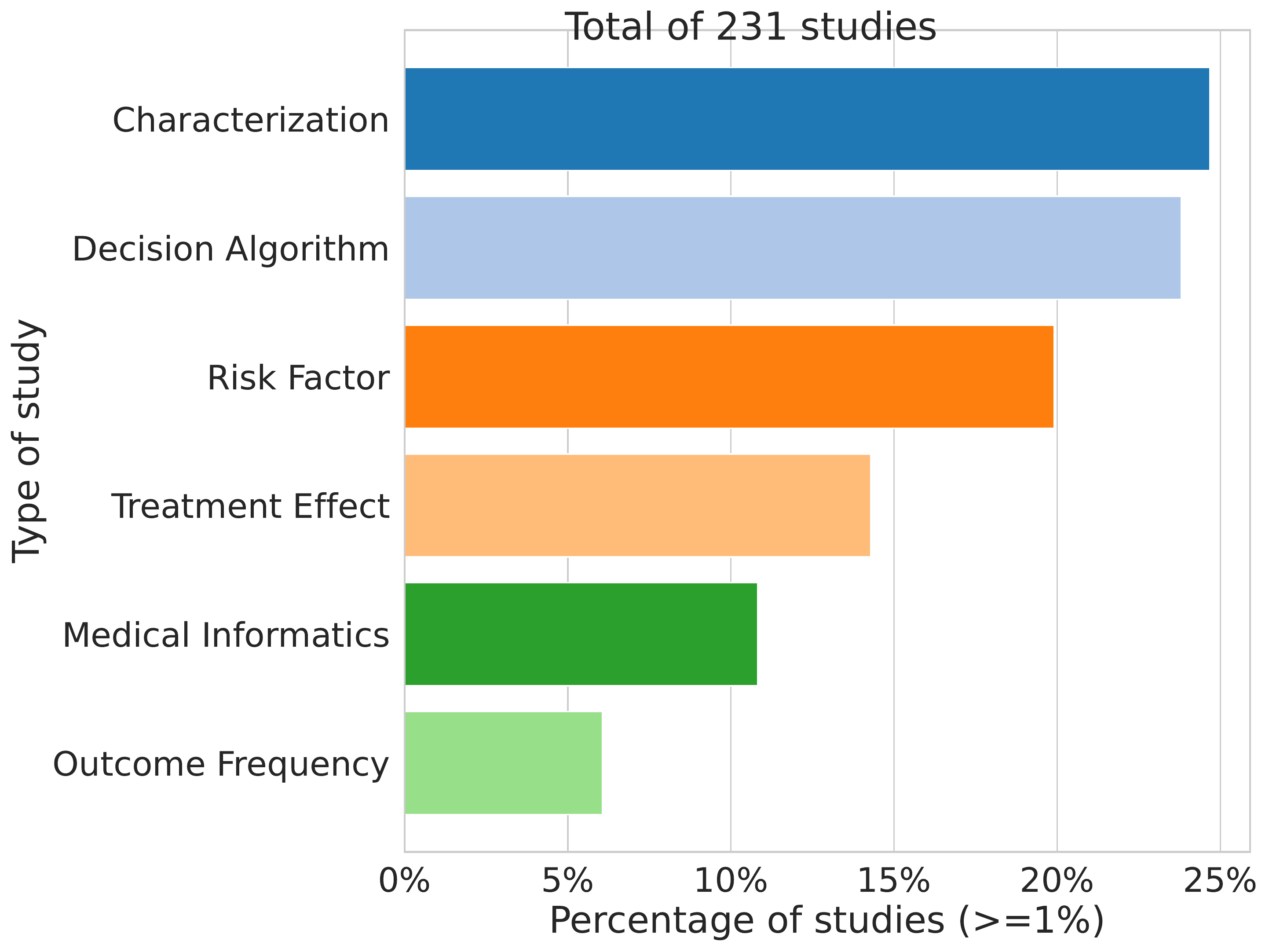}
  \caption{Percentage of studies by objective.}
  \label{results:usage:image:study_objective}
\end{figure*}

The studies are mainly observational (non-interventional). Figure
\ref{results:usage:image:study_objective} presents the distribution of the six
categories defined in \nameref{methods:quantitative} for 231 studies collected
on the study portals of nine hospitals. The studies focus first on population
characterization (25 \%), followed by the development of decision support
processes (24 \%), the study of risk factors (18 \%) and the treatment effect
evaluations (16 \%).

The CDWs are used extensively for internal projects such as student theses (at
least in 9/21) and serve as an infrastructure for single-service research: their
great interest being the de-siloing of different information systems. For most
of the institutions interviewed, there is still a lack of resources and maturity
of methods and tools for conducting inter-institutional research (such as in the
\textit{Grand-Ouest} region of France) or via European calls for projects
(EHDEN). These two research networks are made possible by supra-local governance
and a common data schema, respectively eHop \cite{madec_ehop_2019} and OMOP
\cite{hripcsak_observational_2015}. The Paris hospitals, thanks to its regional
coverage and the choice of OMOP, is also well advanced in multi-centric
research. At the same time, the \textit{Grand-Est} region is building a network
of CDW based on the model of the \textit{Grand-Ouest} region, also using eHop.

\subsubsection*{CDW are used for monitoring and management (16/21)}

The CDW have sometimes been initiated to improve and optimize billing coding
(4/21). The clinical texts gathered in the same database are queried using
keywords to facilitate the structuring of information. The data are then
aggregated into indicators, some of which are reported at the national level.
The construction of indicators from clinical data can also be used for the
administrative management of the institution. Finally, closer to the clinic,
some actors state that the CDW could also be used to provide regular and
appropriate feedback to healthcare professionals on their practices. This
feedback would help to increase the involvement and interest of healthcare
professionals in CDW projects. The CDW is sometimes of interest for health
monitoring (eg. during Covid-19) or pharmacovigilance (13/21).

\subsubsection*{Strong interest for CDW in the context of care (13/21)}

Some CDWs develop specific applications that provide new functionalities
compared to care software. Search engines can be used to query all the
hospital's data gathered in the CDW, without data compartmentalization between
different softwares. Dedicated interfaces can then offer a unified view of the
history of a patient's data, with inter-specialty transversality, which is
particularly valuable in internal medicine. These cross-disciplinary search
tools also enable healthcare professionals to conduct rapid searches in all the
texts, for example to find similar patients \cite{garcelon_finding_2017}. Uses
for prevention, automation of repetitive tasks and care coordination are also
highlighted. Concrete examples are the automatic sorting of hospital
prescriptions by order of complexity, or the setting up of specialized channels
for primary or secondary prevention.

\subsection*{Technical architecture}

The technical architecture of modern CDWs has several layers:
\begin{itemize}
  \item Data processing: connection and export of source data, diverse
        transformation (cleaning, aggregation, filtering, standardization).
  \item Data storage: database engines, file storage (on file servers or object
        storage), indexing engines to optimize certain queries.
  \item Data exposure: raw data, APIs, dashboards, development and analysis
        environments, specific web applications.
\end{itemize}
Supplementary cross-functional components ensure the efficient and secure
operation of the platform: identity and authorization management, activity
logging, automated administration of servers and applications.

The analysis environment (Jupyterhub or RStudio datalabs) is a key component of
the platform, as it allows data to be processed within the CDW infrastructure. A
few CDWs had such operational datalab at the time of our study (6/21) and almost
all of them have decided to provide it to researchers. Currently, clinical
research teams are still often working on data extractions, in less secure
environments.

\subsection*{Data quality, standard formats}

\subsubsection*{Quality tools} Systematic data quality monitoring processes are
being built in some CDWs. Often (8/21), scripts are run at regular intervals to
detect technical anomalies in data flows. Rare data quality investigation tools,
in the form of dashboards, are beginning to be developed internally (3/21).
Theoretical reflections are underway on the possibility of automating data
consistency checks, for example, demographic or temporal. Some facilities
randomly pull records from the EHR to compare them with the information in the
CDW.

\subsubsection*{Standard format}

No single standard data model stands out as being used by all CDWs. All are
aware of the existence of the OMOP (research standard)
\cite{hripcsak_observational_2015} and HL7 FHIR (communication standard) models
\cite{braunstein_health_2019}. Several CDWs consider the OMOP model to be a
central part of the warehouse, particularly for research purposes (9/21). This
tendency has been encouraged by the European call for projects EHDEN, launched
by the OHDSI research consortium, the originator of this data model. In the
\textit{Grand-Ouest} region of France, the CDWs use the eHop warehouse software.
The latter uses a common data model also named eHop. This model will be extended
with the future warehouse network of the  \textit{Grand Est} region also
choosing this solution. Including this grouping and the other establishments
that have chosen eHop, this model includes 12 establishments out of the 32
university hospitals. This allows eHop adopters to launch ambitious
interregional projects. However, eHop does not define a standard nomenclature to
be used in its model and is not aligned with emerging international standards.

\subsubsection*{Documentation}

Half of the CDWs have put in place documentation accessible within the
organization on data flows, the meaning and proper use of qualified data (10/21
mentioned). This documentation is used by the team that develops and maintains
the warehouse. It is also used by users to understand the transformations
performed on the data. However, it is never publicly available. No schema of the
data once it has been transformed and prepared for analysis is published.



\section*{Discussion}\label{discussion}

\subsection*{Principal findings}

We give the first overview of the CDWs in university hospitals of France with 32
hospitals reviewed. The implementation of CDW dates from 2011 and accelerated in
the late 2020. Today, 24 of the university hospitals have an ongoing CDW
project.
From this case study, some general considerations can be drawn, that should be
valuable to all healthcare system implementing CDWs on a national scale.

\subsubsection*{Governance}

As the CDW becomes an essential component of data management in the hospital,
the creation of an autonomous internal team dedicated to data architecture,
process automation and data documentation should be encouraged
\cite{goldacre_better_2022}. This multidisciplinary team should develop an
excellent knowledge of the data collection process and potential reuses in order
to qualify the different flows coming from the source IS, standardize them
towards a homogenous schema and harmonize the semantics. It should have a strong
knowledge of public health competences as well as technical and statistical
competences to develop high quality softwares facilitating the reuses of data.

The resources specific to the warehouse are rare and often taken from other
budgets, or from project-based credits. While this is natural for an initial
prototyping phase, it does not seem adapted to the perennial and transversal
nature of the tool. As a research infrastructure of growing importance, it must
have the financial and organizational means to plan for the long term.

The governance of the CDW has multiple layers: local within the university
hospital, interregional, and national/international. The first level allow to ensure the
quality of data integration as well as the pertinence of data reuse by
clinicians themselves. The interregional level is well adapted for resources
mutualization and collaboration. Finally, the national and international levels
assure coordination, encourage consensus for committing choices such as metadata
or interoperability, and provide financial, technical and regulatory support.

\subsubsection*{Transparency}

International recommendations
\cite{pavlenko_implementation_2020,has_real-world_2021,kohane_what_2021} favour
public referencing of ongoing projects, with prior publication of research
protocols, which is essential from a scientific point of view to control bias.
All institutions should publish all of their studies on
\url{https://clinicaltrials.gov/} in the observational research category.
Introducing EHR as a new subtype of observational study would allow to
better follow the utilization of this emerging data source.

From a patient's perspective, there is currently no way to know if their
personal data is included for a specific project. Better patient information
about the reuse of their data is needed to build trust over the long term. A
strict minimum is the establishment and update of the declarative portals of
ongoing studies for each institution.

\subsubsection*{Data and data usage}

When using CDW, the analyst has not defined the data collection process and is
generally unaware of the context in which the information is logged. This new
dimension of medical research requires a much greater development of data
science skills to change the focus from the implementation of the statistical
design to the data engineering process. Data reuse requires more effort to
prepare the data and document the transformations performed.

International recommendations insist on the need for common data formats
\cite{zhang_best_2022,kohane_what_2021}. However, there is still a lack of
adoption, either of research standards from hospital CDWs to conduct robust
studies with multiple sites, or from EHR vendors to allow sufficient data
interoperability for efficient data communication. Building open-source tools on
top of these standards such as those of OHDSI \cite{schuemie_book_2021} could
foster their adoption.

Many underway studies concern the development of decision support processes
whose goal is to save time for healthcare professionals. These are often
research projects, not yet integrated into routine care. Data reuse oriented
towards primary care is still rare and rarely supported by appropriate funding.

\subsubsection*{Technical architecture}

Tools, methods and data formats of CDW lack harmonization due to the strong
technical innovation and the presence of many actors. As suggested by the recent
report on the use of data for research in the UK \cite{goldacre_better_2022}, it
would be wise to focus on a small number of model technical platforms.

These platforms should favor open source solutions to assure transparency by
default, foster collaboration and consensus and avoid technological lock-in of
the hospitals.

\subsubsection*{Data quality and documentation}

Quality is not sufficiently considered as a relevant scientific topic itself.
However, it is the backbone of all research done within an CDW. In order to
improve the quality of the data with respect to research uses, it is necessary
to conduct continuous studies dedicated to this topic
\cite{zhang_best_2022,kohane_what_2021,shang_conceptual_2018,looten_what_2019}.
These studies should contribute to a reflection on methodologies and standard
tools for data quality, such as those developed by the OHDSI research network
\cite{schuemie_book_2021}.

Finally, there is a need for open source publication of research code to ensure
quality retrospective research
\cite{shang_conceptual_2018,seastedt_global_2022}. Recent research in data
analysis has shown that innumerable biases can lurk in training data sets
\cite{gebru_datasheets_2021,mehrabi_survey_2021}. Open publication of data
schemas is considered an indispensable prerequisite for all data science and
artificial intelligence uses \cite{gebru_datasheets_2021}. Inspired by dataset
cards \cite{gebru_datasheets_2021} and dataset publication guides, it would be
interesting to define a standard CDW card documenting the main data flows.

\subsection*{Limitations}

The interviews were conducted in a semi-structured manner within a limited time
frame. As a result, some topics were covered more quickly and only those
explicitly mentioned by the participants could be recorded. The uneven existence
of study portals introduces a bias in the recording of the types of studies
conducted on CDW. Those with a transparency portal already have more maturity in
use cases.

With only one oncology specialized center and four non-university
hospital groups, including two private health care institutions, we have not
covered the exhaustive health care landscape in France. CDW initiatives also
exist in primary care, in smaller hospital groups and in private companies.

\section*{Conclusion}\label{conclusion}
The French CDW ecosystem is beginning to take shape, benefiting
from an acceleration thanks to national funding, the multiplication of
industrial players specializing in health data and the beginning of a
supra-national reflection on the
European
Health Data Space\cite{ehds_2022}. However, some points require special attention to ensure
that the potential of the CDW translates into patient benefits.

The priority is the creation and perpetuation of multidisciplinary warehouse
teams capable of operating the CDW and supporting the various projects. A
combination of public health, data engineering, data stewardship, statistics and
IT competences is a prerequisite for the success of the CDW. The team should be
the privileged point of contact for data exploitation issues and should
collaborate closely with the existing hospital departments.

The constitution of a multi-level collaboration network is another priority. The
local level is essential to structure the data and understand its possible uses.
Interregional, national and international coordination would make it possible to
create thematic working groups, in order to stimulate a dynamic of cooperation
and mutualization.

A common data model should be encouraged, with precise metadata allowing to map
the integrated data, in order to qualify the uses to be developed today from the
CDWs. More broadly, open-source documentation of data flows and transformations
performed for quality enhancement would require more incentives to unleash the
potential for innovation for all health data reusers.

Finally, the question of expanding the scope of the data beyond the purely
hospital domain must be asked. Many risk factors and patient follow-up data are
missing from the CDWs, but are crucial for understanding pathologies. Combining
city data and hospital data would provide a complete view of patient care.

\section*{Ethics Statement}

This work has been authorized by the board of the French Health Authority of
Health (HAS). Every interviewed participant was asked by email for their participation and
informed on the possible forms of publication: a French official report and
international publication. Furthermore, at each interview, every participant has
been asked for their agreement before recording the interview. Only one
participant refused the video to be recorded.

\section*{Acknowledgments}
\subsection*{Funding}

The Haute Autorité de Santé (HAS) funded this research; The Inria supervised
Matthieu Doutreligne in the Social Data team.

\subsection*{Author contributions}

Conceptualization: PAJ, MD
Data curation: MD
Formal analysis: MD
Methodology: MD, AD
Project Administration: PAJ
Software: MD
Writing - Original Draft Preparation: md
Writing - Review \& Editing: MD, AD, PJ, AL, XT

\subsection*{Acknowledgments}

We want to thanks all participants and experts interviewed for this study. We
also want to thanks other people that proof read the manuscript for external
review : Judith Fernandez (HAS), Pierre Liot (HAS), Bastien Guerry (Etalab),
Aude-Marie Lalanne Berdouticq (Institut Santé numérique en Société), Albane
Miron de L’Espinay (ministère de la Santé et de la Prévention), Caroline Aguado
(ministère de la Santé et de la Prévention). We also thank Gaël
Varoquaux for his support and advice.


%
%

\printbibliography

\clearpage
\appendix
\section*{Appendices}
\section{List of interviewed stakeholders with their
  teams}\label{apd:table:expert_teams}

\begin{table}[!ht]
  \centering
  \begin{tabular}{ll}
    \thickhline
    Clinical Data Warehouse & Teams                                             \\
    \thickhline
    CDW\_AMIENS             & IT : 1,MID : 1                                    \\
    CDW\_ANGERS             & Data Direction : 1                                \\
    CDW\_APHM               & Clinician : 1,CDW team : 2                        \\
    CDW\_APHP               & CDW team : 4,IT : 5                               \\
    CDW\_BORDEAUX           & CDW team : 1,Inserm : 1,public health : 2         \\
    CDW\_BREST              & CDW team : 1,MID : 1                              \\
    CDW\_DIJON              & CDW team : 1                                      \\
    CDW\_EDSAN              & CDW team : 2,MID : 1                              \\
    CDW\_HCL                & Clinician : 1,Data Direction : 1,IT : 1,Inserm    \\
    CDW\_INCLUDE\_LILLE     & Administration : 2,CDW team : 3,public health : 2 \\
    CDW\_MARTINIQUE         & CDW team : 1,public health : 1                    \\
    CDW\_MONTPELLIER        & Data Direction : 2,MID : 1,public health : 1      \\
    CDW\_NANCY              & CDW team : 2,public health : 2                    \\
    CDW\_NANTES             & CDW team : 2,public health : 1                    \\
    CDW\_POITIERS           & IT : 2,CRD : 1                                    \\
    CDW\_PREDIMED\_CHUGA    & CDW team : 3,public health : 2                    \\
    CDW\_REIMS              & Clinician : 1,CDW team : 1                        \\
    CDW\_RENNES             & CDW team : 2,public health : 2                    \\
    CDW\_STRASBOURG         & CDW team : 2,public health : 2                    \\
    CDW\_TOULOUSE           & CDW team : 1                                      \\
    CDW\_TOURS              & CDW team : 1                                      \\
    \thickhline
  \end{tabular}
\end{table}

\clearpage

\begin{landscape}
  \section{Interview form} \label{apd:interview_form}
  \begin{table}[h!]
    \centering
    \resizebox{1.3\textwidth}{!}{%
      \begin{tabular}{|l|l|}
        \hline
        \textbf{Topics}                                                 &
        \textbf{Questions}
        \\ \hline
        \multirow{5}{*}{\begin{tabular}[c]{@{}l@{}}Initiation and Construction
            of \\ the Clinical Data Warehouse\end{tabular}}                     &
        How was the initiative born, when, which team(s) involved in the
        construction? A Data warehouse to meet what initial needs ?                                                          \\
        \cline{2-2}                                                     & What was (is) the articulation between the medical
        informatics / engineer(s) / Clinical Research Department  and user
        team(s), biostatistics ?                                                                                             \\ \cline{2-2} & Governance: How
        should the teams be organized for the creation and maintenance of the
        warehouse, data access, and project teams?                                                                           \\ \cline{2-2} &
        \begin{tabular}[c]{@{}l@{}}What types of data are present in the
          warehouse
          from the
          following
          non-exhaustive
          list: \\
          Billing
          codes ,
          other
          administrative
          data,
          other
          procedures,
          structured
          procedures
          and
          diagnoses,
          structured
          biology
          measures,
          \\
          structured
          drug
          treatments,
          emergencies,
          resuscitation,
          anesthesia,
          texts
          (letters,
          Clinician
          Reports),
          imaging,
          anatomopathology,
          sequencing.\end{tabular}
        \\
        \cline{2-2}
                                                                        & What
        are the
        medico-social/social
        data,
        especially
        from
        social
        and
        medico-social
        institutions
        ?
        \\
        \hline
        \multirow{6}{*}{Current status - Ongoing and finished projects} &
        \begin{tabular}[c]{@{}l@{}}Who are the main users? For what purposes
          (research,
          quality
          improvement,
          management,
          clinical
          usage)?
          \\ Which
          therapeutic
          area(s)?\end{tabular}
        \\
        \cline{2-2}
                                                                        &
        \begin{tabular}[c]{@{}l@{}}What are the major types of projects from the
          following
          non-exhaustive
          list: \\
          Cohort
          development,
          descriptive
          epidemiology,
          analytical
          (comparative)
          epidemiology
          with/without
          randomization,
          monitoring
          and
          dashboards,
          \\indicators,
          inclusion
          in
          clinical
          trials.\end{tabular}
        \\
        \cline{2-2}
                                                                        & How
        many
        projects
        are
        completed
        /
        started
        /
        planned?
        \\
        \cline{2-2}
                                                                        & What
        are the
        tools
        and
        methods
        used for
        these
        projects?
        Cohort
        building
        tool,
        standard
        data
        formats,
        NLPs,
        ...
        \\
        \cline{2-2}
                                                                        & Is
        there a
        valorization
        strategy
        for the
        Clinical
        Data
        Warehouse?
        \\
        \cline{2-2}
                                                                        & What
        connections
        with
        external
        sources
        such as
        the
        national
        health
        data
        platform,
        the
        outpatient
        data,
        the
        general
        practitioner
        data,
        the
        research
        cohorts
        ?                                                                                                                    \\
        \hline
        \multirow{3}{*}{Opportunity and obstacles}                      & What
        are the main difficulties encountered during data warehouse projects?
        \\ \cline{2-2} & Are there themes that deserve more encouragement from
        the HAS?
        \\ \cline{2-2} & What skills are needed? Are there any skills or
        technical resources missing?
        \\ \hline
        \multirow{11}{*}{\begin{tabular}[c]{@{}l@{}}Quality criteria for
            observational research\end{tabular}}                    & Coverage: How
        is it monitored? Geographically/by department? Time-wise? By what means?
        \\ \cline{2-2} & Cleaning: How are patient duplicates and source
        alignment managed?
        \\ \cline{2-2} &
        \begin{tabular}[c]{@{}l@{}}Database Network: Does the warehouse belong
          to a
          health
          database
          network?\end{tabular}
        \\
        \cline{2-2}
                                                                        &
        \begin{tabular}[c]{@{}l@{}}Data quality: Are there automatic reports on
          data
          quality?
          Frequency,
          design,
          code and
          documentation
          available?
          Presence
          of
          dedicated
          personnel
          \\ or
          even a
          team to
          check
          the
          quality
          of the
          data
          continuously,
          and to
          carry
          out
          quality
          controls
          of the
          data on
          the
          central
          base, on
          the
          study
          bases?\end{tabular}
        \\
        \cline{2-2}
                                                                        &
        \begin{tabular}[c]{@{}l@{}}Data life cycle: Is there a reference
          document
          on the
          different
          stages
          of the
          data
          life
          cycle?
          \\ How
          is this
          document
          kept up
          to date
          with the
          constant
          evolution
          of the
          warehouse?
          In what
          form? \\
          How is
          this
          documentation
          managed,
          accessed,
          updated
          and
          corrected?
          Precise
          description
          of the
          integrated
          fields?\end{tabular}
        \\
        \cline{2-2}
                                                                        &
        \begin{tabular}[c]{@{}l@{}}Harmonization procedure: What are the data
          structures/formats
          and
          coding
          systems
          used? \\
          (eHop,
          I2B2,
          OMOP,
          HL7
          FHIR,
          other
          ?)\end{tabular}
        \\
        \cline{2-2}
                                                                        &
        \begin{tabular}[c]{@{}l@{}}Machine learning: If machine learning systems
          are used
          (e.g.
          for
          extracting
          and
          structuring
          information),
          is there
          specific
          documentation
          on their
          performance?
          \\ For
          manual
          coding
          (e.g.
          labelling),
          is there
          a coding
          guide?
          Has a
          measurement
          of
          inter-coder
          consistency
          been
          conducted?\end{tabular}
        \\
        \cline{2-2}
                                                                        &
        De-identification:
        Elements
        on
        de-identification
        if
        applicable,
        performance
        metrics
        \\
        \cline{2-2}
                                                                        &
        \begin{tabular}[c]{@{}l@{}}Constructed phenotypes: Are there operational
          definitions
          of
          target
          populations
          (study
          cohorts)
          and how
          are
          these
          compared
          to
          conceptual
          definitions,
          i.e., \\
          business
          and
          scientific
          definitions?
          Is there
          a study
          of
          FPR/TPR
          in
          relation
          to a
          reference
          standard?
          Are
          these
          definitions
          made
          public
          either
          with the
          study \\
          results
          or in
          the
          documentation
          of the
          warehouse?\end{tabular}
        \\
        \cline{2-2}
                                                                        &
        \begin{tabular}[c]{@{}l@{}}Transparency: Are the studies registered on a
          dedicated
          or
          pre-existing
          portal
          (epidemio-France,
          encepp
          (EU),
          clinicaltrials.gov
          (US))?
          \\ Are
          the
          study
          codes
          made
          accessible
          as for
          opensafely?
          Are the
          publications
          accessible
          in open
          access,
          once the
          studies
          are
          completed?\end{tabular}
        \\
        \cline{2-2}
                                                                        &
        \begin{tabular}[c]{@{}l@{}}Multidisciplinarity: Are the project teams
          multidisciplinary? Specification of the participations for each part of
          the analysis from the data collection \\ from the raw Information
          System.\end{tabular}
        \\ \hline
        \multirow{2}{*}{Topics of interest to the HAS}                  &
        \begin{tabular}[c]{@{}l@{}}Quality Department: quality indicators
          (french
          IQSS):
          coordination
          (patient
          assessment
          for
          discharge,
          patient
          contact
          at D+1),
          quality
          of the
          liaison
          letter),
          \\
          management
          (eligibility
          for the
          outpatient
          surgeries,
          pain
          management)\end{tabular}
        \\
        \cline{2-2}
                                                                        &
        \begin{tabular}[c]{@{}l@{}}Health Technology Assessment Department:
          hospital biology activity (description), adverse events associated with
          procedures, post-registration studies \\(procedures, early access
          oncology). Evaluation of procedures: e.g. biological and imaging
          procedures performed in hospitals, genetic tests in oncology and rare
          diseases.\end{tabular}
        \\ \hline
        open discussion                                                 &                                                    \\
        \hline
      \end{tabular}%
    }

  \end{table}
\end{landscape}

\section{Study data tables}\label{apd:study_tables}

The data tables used to produce the figures in the results section are available
at the following url:

\url{https://gitlab.has-sante.fr/has-sante/public/rapport_edsh/}.

The guests table concerns the individuals interviewed, the interview dates, the
positions and the membership of a specific team. The warehouse table collects
information about the CDW. The table of studies is the referencing of the
studies informed on 10 portals of studies in progress (or completed if
available) available in free access.

\end{document}